\documentclass[12pt]{article}
\let\section=\subsection     \let\subsection=\subsubsection

\usepackage{graphicx}

\begin{document}
\begin{center}
   {\large \bf THERMAL MODEL AT RHIC:}\\[2mm]
   {\large \bf PARTICLE RATIOS AND $p_\perp$ SPECTRA}\\[5mm]
   Wojciech BRONIOWSKI and Wojciech FLORKOWSKI \\[5mm]
   {\small \it The H. Niewodnicza\'nski Institute of Nuclear Physics \\
   PL-31342 Cracow, POLAND \\[8mm] }
\end{center}

\begin{abstract}\noindent
Predictions of the single-freeze-out model for the particle
spectra at RHIC are presented. The model assumes that the 
chemical and thermal freeze-outs occur simultaneously, and 
incorporates in simple terms the 
longitudinal and transverse flow. All resonance decays 
are included. The model predictions 
and the data are in striking agreement in the whole available 
range of momenta.
\end{abstract}

Much has been said during this Workshop on the predictions for the particle
ratios within the thermal approach (see the contribution by J. Stachel),
hence in this talk we concentrate
entirely on the $p_\perp$ spectra \cite{wbwf,wbwfs}. 
We only wish to mention that
our study \cite{wfwbmm} of the
particle ratios at RHIC
confirms, within statistical errors, 
the results of \cite{pbmrhic}, with the following values of the
thermal parameters:
\begin{equation}
T=165 \pm 7 \hbox{ MeV}, \,\,\mu_{B}=41 \pm 5 \hbox{ MeV}.
\label{thermo}
\end{equation}
Strangeness conservation gives $\mu_S=9$ MeV, and isospin asymmetry
of the gold nuclei yields $\mu_I=-1$ MeV. For more details, review of the
thermal models, and references, see \cite{mm}.

Our model for the spectra incorporates the following assumptions:

\begin{enumerate}

\item Chemical and thermal freeze-outs occur {\em simultaneously} 
on a freeze-out hypersurface.
In other words, elastic rescattering after the freeze-out is neglected.
This simplification, which opposes the traditional picture with 
two distinct freeze-outs at noticeably different temperatures, 
works very well, as can be seen from our figures.
Note that rapid expansion of the system inhibits the collision rate.
Also, the Van der Waals corrections discussed at the end of this talk
make the system more dilute at freeze-out, reducing its opacity.
Arguments hinting rapid expansion have also been presented in the 
contributions by H. Appelsh\"auser and T. Hirano.   

\item At freeze-out the particles are distributed according to the 
statistical distribution functions. The two thermodynamic 
parameters (\ref{thermo})
are obtained from the analysis of the ratios of the particle multiplicities.
Since these depend rather weakly on centrality, we treat the thermal
parameters as universal.

\item The resonances are treated in a {\em complete} 
way, with all particles from
the PDG tables incorporated. Their sequential decays are included exactly,
in a semi-analytic fashion. The role of the resonances is very important,
resulting in substantial ``cooling'' of the spectra \cite{wfwbmm}.

\item The freeze-out hypersurface is a simple generalization of the Bjorken
model. It
is parameterized with two geometric parameters specified below:
the invariant time, $\tau$, and the transverse radius, $\rho_{\rm max}$.
For simplicity, Hubble-like expansion is assumed, with the four-velocity
proportional to the coordinate, $u^\mu = x^\mu/\tau$. Clearly, other
parameterizations of the fire-ball expansion,
{\em e.g.} those following from the hydrodynamic evolution in various
models, can be implemented and tested.

\item The {\em only} parameters of the model are the two universal thermal
parameters (\ref{thermo})
and the two geometric parameters, $\tau$ and $\rho_{\rm max}$
(different for each centrality).

\end{enumerate}

The parameterization of the freeze-out hypersurface, taken
in the spirit of the Buda-Lund model \cite{buda}, is as follows:
$$t = {\rm ch} \alpha _{\parallel} \sqrt{\tau^2+\rho^2},
\quad {z}={\rm sh} \alpha _{\parallel}  \sqrt{\tau^2+\rho^2}, \quad
{x} = \rho \cos \phi ,\quad {y}=\rho \sin \phi,$$ 
with $\rho \le \rho_{\rm max}$. 
The parameter $\tau$ fixes the
overall normalization of the spectra, while the ratio 
 $\rho_{\rm max}/\tau$ determines their shape.
The standard Cooper-Frye formalism is applied to obtain the
spectra. The details  of our procedure, including the technicalities of the
resonance decays, are given in the Appendix of Ref. \cite{wbwfs}.

Fig. 1 shows our results for the
PHENIX minimum-bias data at $\sqrt{s}=130$ GeV A \cite{velko}. The
fitted values of the geometric parameters are
$\tau = 5.55$ fm and $\rho_{\rm max} = 4.50$ fm. We note
excellent agreement in the whole available data range, with the 
magnitude of the spectra 
spanned over five decades. Virtually all 
points are crossed by the model curves within error bars. Similar agreement,
not shown, has been 
found for $\pi^0$. Note that all the non-trivial experimental features 
are reproduced in Fig. 1. In particular, we find the convex shape of the pion
spectra, the crossing of $p$ and $\pi^+$ around $p_\perp = 2$ GeV, and of 
$K^+$ and 
\begin{center}
   \includegraphics[width=14cm]{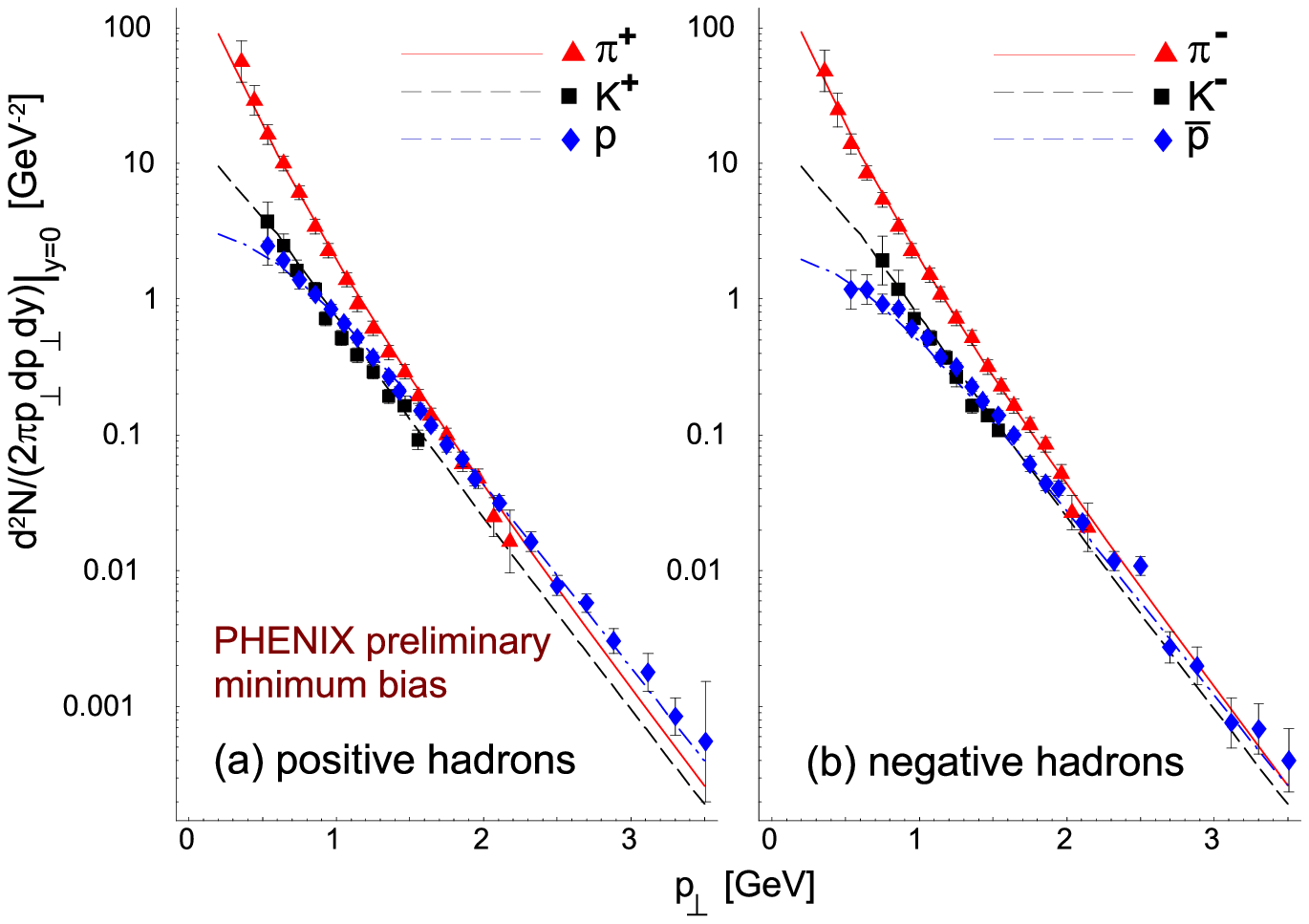}\\
   \parbox{14cm}
        {{\footnotesize 
        Fig.~1: Predictions of the single-freeze-out model for the
particle spectra at mid-rapidity vs. the PHENIX 
minimum-bias data.}}
\end{center}
\noindent
$\pi^+$ around 1 GeV (we note in passing that a similar effect of
crossing occurs for the SPS data plotted in the $p_\perp$ variable, 
{\em cf.} Fig. 3). The model results at $p_\perp >2$ GeV should be taken 
with a grain of salt, since hard processes are expected to be important
in that region (see the talk by P. Levai).  

Fig. 2 shows our results for the most central collisions at RHIC, where we 
find from the fit
$\tau = 7.66$ fm and $\rho_{\rm max} = 6.69$ fm \cite{wbwf}. 
These numbers are,
as expected from geometry,  
larger than for the minimum-bias case, which averages over centralities. 
The data in Fig. 2 come from
\cite{velko,harris,yama,snellings}. 
All data except for $\overline{\Lambda}$ are absolutely normalized.
The normalization for $\overline{\Lambda}$, not available experimentally in 
\cite{snellings}, 
has been adjusted arbitrarily. We have found later that this norm
is consistent with the newly-released normalized data of \cite{bel}.  
The curves for other hyperons are predictions.
Again, the agreement in Fig. 2 is remarkable, 
except for the $\overline{p}$ data from 
STAR, however these data have been corrected for weak decays through the use 
of the HIJING model \cite{harris}, which led to about 20\% decrease. 
Our model curves in all plots
include the full feeding from the weak decays. 
The fact that our approach reproduces the $\overline{p}$ yields brings a 
support for the single-freeze-out idea. This is because the
large annihilation cross section of $\overline{p}$ would decrease its 
abundance if it 
\begin{center}
   \includegraphics[width=14cm]{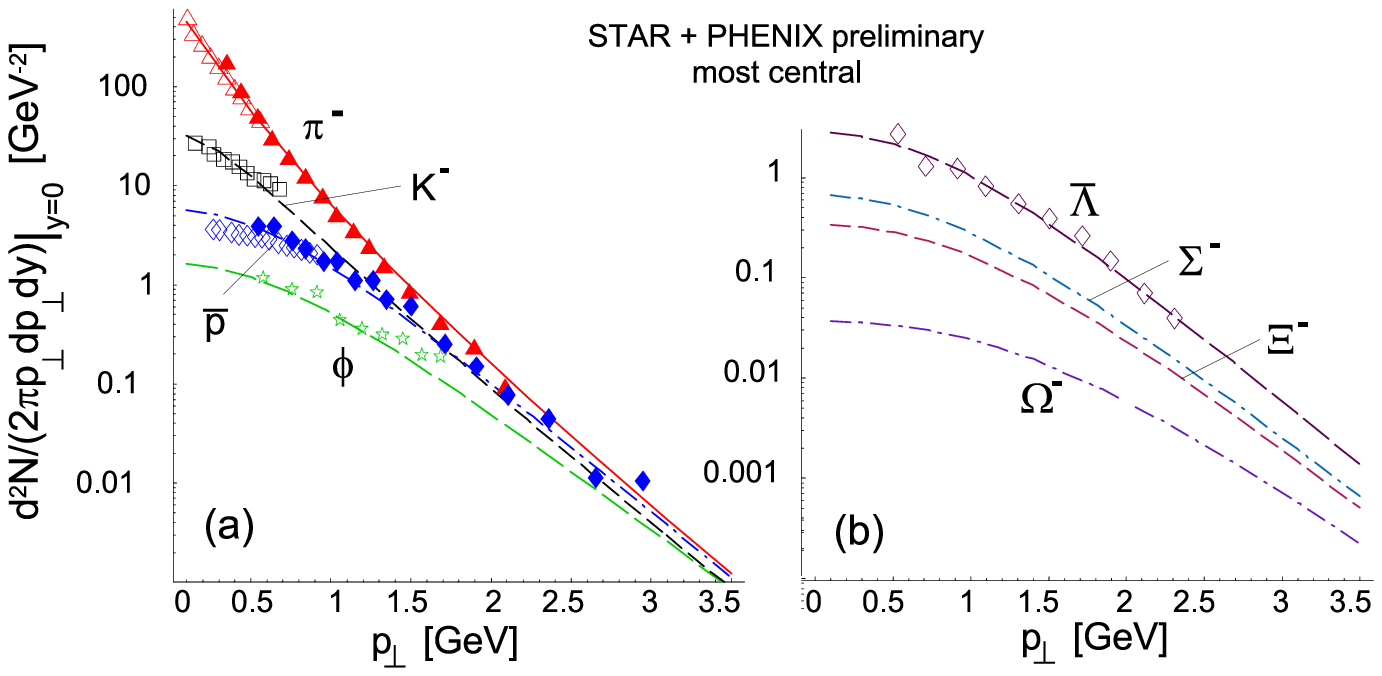}\\
   \parbox{14cm}
        {{\footnotesize 
        Fig.~2: Same as Fig. 1 for the most-central data from STAR (open
symbols) and PHENIX (filled symbols) .}}
\end{center}
\noindent
interacted with the protons. Thus the single-freeze-out
solves the anti-baryon puzzle discussed in the talk by R. Rapp.
The fact that the $\phi$ data are well reproduced also deserves a credit.
This meson interacts rather weakly with the medium, thus it can serve as
an accurate thermometer of the system at freeze-out. 
Fits of similar quality as in Fig. 2 can be found 
for non-central collisions.

The value of  $\rho_{\rm max}$ of Fig. 2 
leads to the following values of the 
average and maximum transverse flow velocity: 
$\langle \beta_\perp \rangle = 0.49$, $\beta^{\rm max}_\perp = 0.66$.

Fig. 3 shows the model fits to the most central NA49 data. For all 
particles very reasonable agreement is achieved, 
except for the $\Omega^-$ (not shown), where the model slope is much too
low.

Now we remark on the problem of the HBT radii. As pointed out in the talk by 
D. Hardtke, our transverse size $\rho_{\rm max}$ is small, such that 
the expected value of $R_s$ would be too low, of the order of 3.5 fm.
However, there are two 
important effect which increase $R_s$. The first is the decay of resonances. 
The resonances travel about of 1 fm before they decay, which
increases the radius. Since three quarters of pions come
from resonance decays, this increases $R_s$. The other, more pronounced, 
effect is the excluded volume/Van der Waals 
correction. The excluded volume correction brings in a factor of 
$(1+ v n)^{-1}$, 
where $v \simeq 5$ fm$^3$ is the eigenvolume of the particle 
(assumed, for simplicity, the same for all particles) 
and $n \simeq 0.5 {\rm fm}^{-3}$ is the density
of particles. Another factor comes from the modification of 
the chemical potential, and 
(for the Boltzmann distributions) has the form $\exp(-pv/T)$, where 
$p$ is the pressure. 
\begin{center}
   \includegraphics[width=14cm]{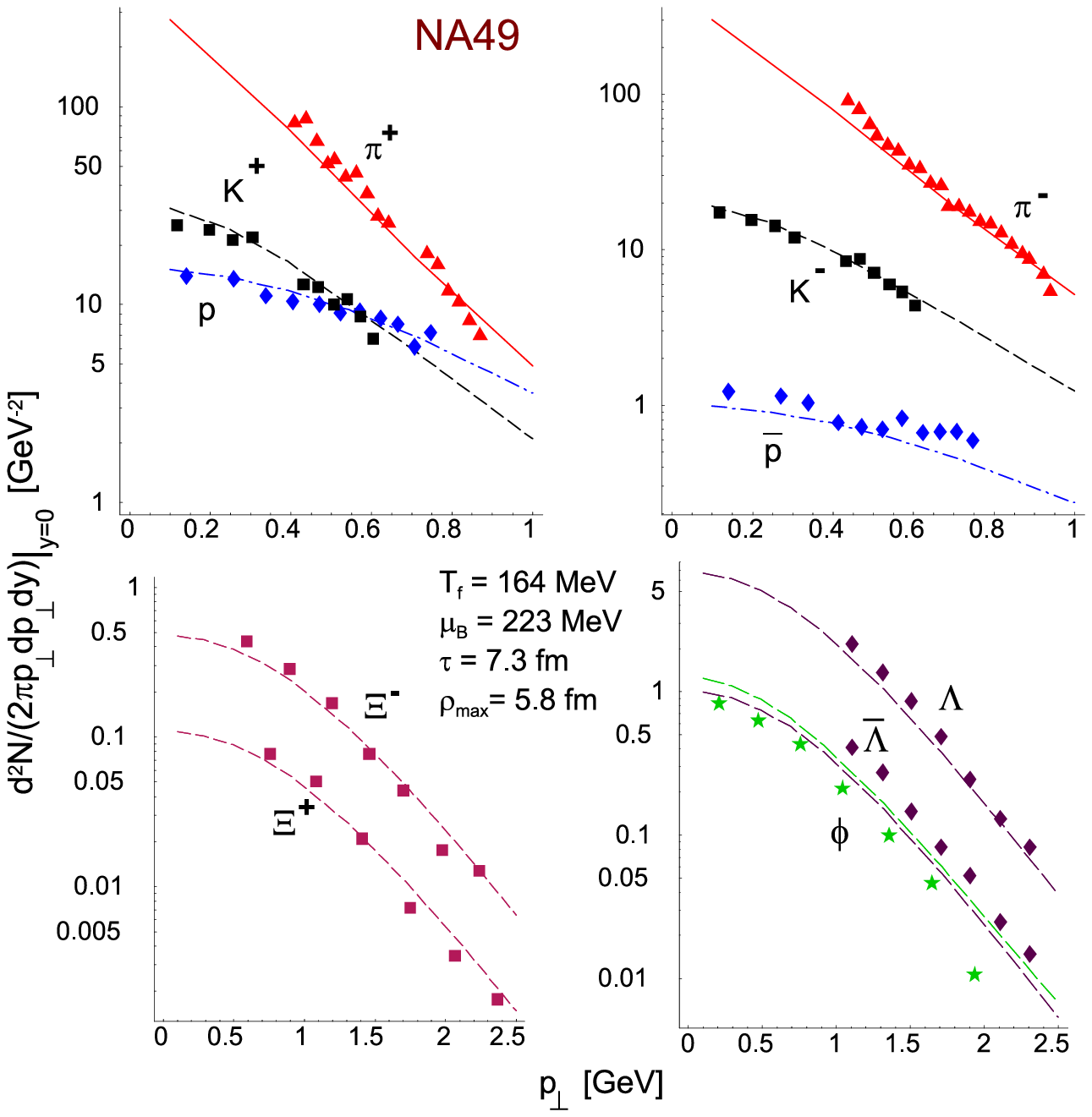}\\
   \parbox{14cm}
        {{\footnotesize 
        Fig.~3: Same as Fig. 1 for the SPS NA49 data 
(Nucl. Phys. {\bf A610} (1996) 188c; Phys. Lett. {\bf B444} (1998) 523; 
Phys. Lett. {\bf B491} (2000) 59).}}
\end{center}
\noindent
The product
of these factors, which we denote by $s$, 
is significantly smaller than 1.
It changes the normalization of the 
integrals for the particle multiplicities and 
spectra. It can be compensated by 
simultaneously rescaling $\tau$ and $\rho_{\rm max}$ by
$s^{-1/3}$. The procedure leaves the particle ratios and the spectra intact (!),
however, it scales up the HBT radii by $s^{-1/3}$. A detailed 
self-consistent inclusion of the Van der Waals corrections is necessary 
to obtain detailed estimates. We expect that the needed 50-60\%
increase will follow naturally. Hence, our approach has no 
fundamental problem with the size of the HBT radii,
provided the resonance decays and the 
Van der Waals effect is incorporated. Also, with the 
inclusion of the Van der Waals effects the size parameters
are raised comfortably above the geometric size of the gold nuclei, making 
room and time for the collective flow to develop.
Certainly, it is important to see in detail what are the predictions of our 
model for all HBT radii, in particular for the ratios, where most 
approaches have serious problems. This research is under way.

In conclusion, the agreement with the 
$p_\perp$ spectra for all
species of particles, achieved 
with just two thermal and two geometric parameters, is surprisingly good 
{\em cf.} Figs. 1-2.  Thus, our analysis
provides a very strong support for the thermal approach
to the particle production at RHIC. 

\bigskip

We are very grateful to Marek Ga\'zdzicki for numerous 
clarifying discussions.
This work has been supported by the Polish State Committee for
Scientific Research, grant 2 P03B 09419.

\end{document}